\documentclass{article}

\usepackage{arxiv}

\usepackage[utf8]{inputenc}
\usepackage[T1]{fontenc}

\usepackage{microtype}
\usepackage{booktabs}
\usepackage{amsfonts}
\usepackage{nicefrac}
\usepackage{float}
\usepackage{cuted}
\usepackage{enumitem}

\usepackage[numbers,sort&compress]{natbib}
\usepackage{doi}                 
\usepackage{graphicx}
\usepackage{url}

\usepackage{flushend}

\usepackage[acronym,nonumberlist]{glossaries}

\newacronym{sota}{SOTA}{state-of-the-art}
\newacronym{dpm}{DPM}{dynamic process management}
\newacronym{slurm}{SLURM}{Simple linux utility for resource management}
\newacronym{mqss}{MQSS}{Munich quantum software stack}
\newacronym{qdmi}{QDMI}{quantum device management interface} 
\newacronym{hcqw}{HCQW}{hybrid classical-quantum workflow} 
\newacronym{hpc}{HPC}{high performance computing}
\newacronym{wlm}{WLM}{workload manager}
\newacronym{qec}{QEC}{quantum error correction}
\newacronym{qem}{QEM}{quantum error mitigation}
\newacronym{nisq}{NISQ}{noisy, intermediate-scale quantum} 
\newacronym{mpi}{MPI}{message passing interface}
\newacronym{api}{API}{application programming interface}
\newacronym{rpc}{RPC}{remote procedure call}
\newacronym{hhl}{HHL}{Harrow-Hassidim-Lloyd}
\newacronym{qpu}{QPU}{quantum processing unit}
\newacronym{cpu}{CPU}{central processing unit}
\newacronym{gpu}{GPU}{graphics processing unit}
\newacronym{fpga}{FPGA}{field-programmable gate array}
\newacronym{ghz}{GHZ}{Greenberger-Horne-Zeilinger}
\newacronym{mpmd}{MPMD}{multiple program, multiple data}
\newacronym{qasm}{QASM}{quantum assembly language}
\newacronym{isc}{ISC}{International Supercomputing Conference}
\newacronym{mtbf}{MTBF}{mean time between failures}
\makeglossaries

\usepackage{xcolor}
\usepackage{listings}

\definecolor{lightgray}{gray}{0.97}
\definecolor{darkgray}{gray}{0.35}
\definecolor{purple}{rgb}{0.58,0,0.82}
\definecolor{bluekeywords}{rgb}{0.13,0.13,1.0}
\definecolor{tealstrings}{rgb}{0.0,0.5,0.5}

\lstdefinelanguage{LLVM}{
  morekeywords={declare,define,call,ret,br,phi,icmp,bitcast,label,type,opaque,void},
  sensitive=true,
  morecomment=[l]{;},
  morestring=[b]",
}

\title{A Full Stack Framework for High Performance Quantum-Classical Computing}

\author{ {Xin Zhan}\thanks{Corresponding author.} \\
	Hewlett Packard Enterprise\\
	Houston, USA \\
	\texttt{xin.zhan@hpe.com} \\
	\And
	{\hspace{1mm}K. Grace Johnson} \\
	Hewlett Packard Enterprise\\
	Milpitas, USA \\
	\texttt{grace.johnson@hpe.com} \\
    \And
	{\hspace{1mm}Aniello Esposito} \\
	Hewlett Packard Enterprise\\
	Basel, Switzerland\\
	\texttt{aniello.esposito@hpe.com} \\
    \And
	{\hspace{1mm}Barbara Chapman} \\
	Hewlett Packard Enterprise\\
	New York, USA\\
	\texttt{barbara.chapman@hpe.com} \\
    \And
	{\hspace{1mm}Marco Fiorentino} \\
	Hewlett Packard Enterprise\\
	Milpitas, USA\\
	\texttt{marco.fiorentino@hpe.com} \\
    \And
	{\hspace{1mm}Kirk M. Bresniker} \\
	Hewlett Packard Enterprise\\
	Milpitas, USA\\
	\texttt{kirk.bresniker@hpe.com} \\
    \And
	{\hspace{1mm}Raymond G. Beausoleil} \\
	Hewlett Packard Enterprise\\
	Milpitas, USA\\
	\texttt{ray.beausoleil@hpe.com} \\
    \And
	{\hspace{1mm}Masoud Mohseni} \\
	Hewlett Packard Enterprise\\
	Milpitas, USA\\
	\texttt{masoud.mohseni@hpe.com} \\ 
}

\date{}

\renewcommand{\headeright}{}
\renewcommand{\undertitle}{}

\hypersetup{
pdfauthor={Xin Zhan},
pdfkeywords={},
}
\begin{document}
\maketitle

\begin{abstract}
To address the growing needs for scalable High Performance Computing (HPC) and Quantum Computing (QC) integration, we present our HPC-QC full stack framework and its hybrid workload development capability with modular hardware/device-agnostic software integration approach. The latest development in extensible interfaces for quantum programming, dispatching, and compilation within existing mature HPC programming environment are demonstrated. Our HPC-QC full stack enables high-level, portable invocation of quantum kernels from commercial quantum SDKs within HPC meta-program in compiled languages (C/C++ and Fortran) as well as Python through a quantum programming interface library extension. An adaptive circuit knitting hypervisor is being developed to partition large quantum circuits into sub-circuits that fit on smaller noisy quantum devices and classical simulators. At the lower-level, we leverage Cray LLVM-based compilation framework to transform and consume LLVM IR and Quantum IR (QIR) from commercial quantum software front-ends in a retargetable fashion to different hardware architectures. Several hybrid HPC-QC multi-node multi-CPU and GPU workloads (including solving linear system of equations, quantum optimization, and simulating quantum phase transitions) have been demonstrated on HPE EX supercomputers to illustrate functionality and execution viability for all three components developed so far. This work provides the framework for a unified quantum-classical programming environment built upon classical HPC software stack (compilers, libraries, parallel runtime and process scheduling).
\end{abstract}


\pagebreak
\twocolumn
\section{Introduction}

%
A hybrid HPC-QC computation paradigm, where quantum processing units (QPUs) are integrated into heterogeneous HPC infrastructures as additional accelerators besides GPUs and FPGAs, will maximize the optimal utilization of both paradigms for massive parallel processing and enable quantum computing to operate at scale \cite{ref1}. The synergy between HPC and QC provides a unique advantage to tackle challenges such as complex memory requirements for QC and computationally intractable problems in HPC that neither paradigm can efficiently solve on its own. Due to potential exponential reductions in time or space complexity for certain problems, QC has attracted significant interest as an alternative computational model for performance beyond exascale. The desire for HPC-QC hybrid systems provides natural and
essential demands for integrating quantum programming capability (to program, compile, distribute, and execute quantum circuits) into existing classical HPC programming environments.

\begin{figure}[!htbp]
	\centering
	\includegraphics[width=0.48\textwidth]{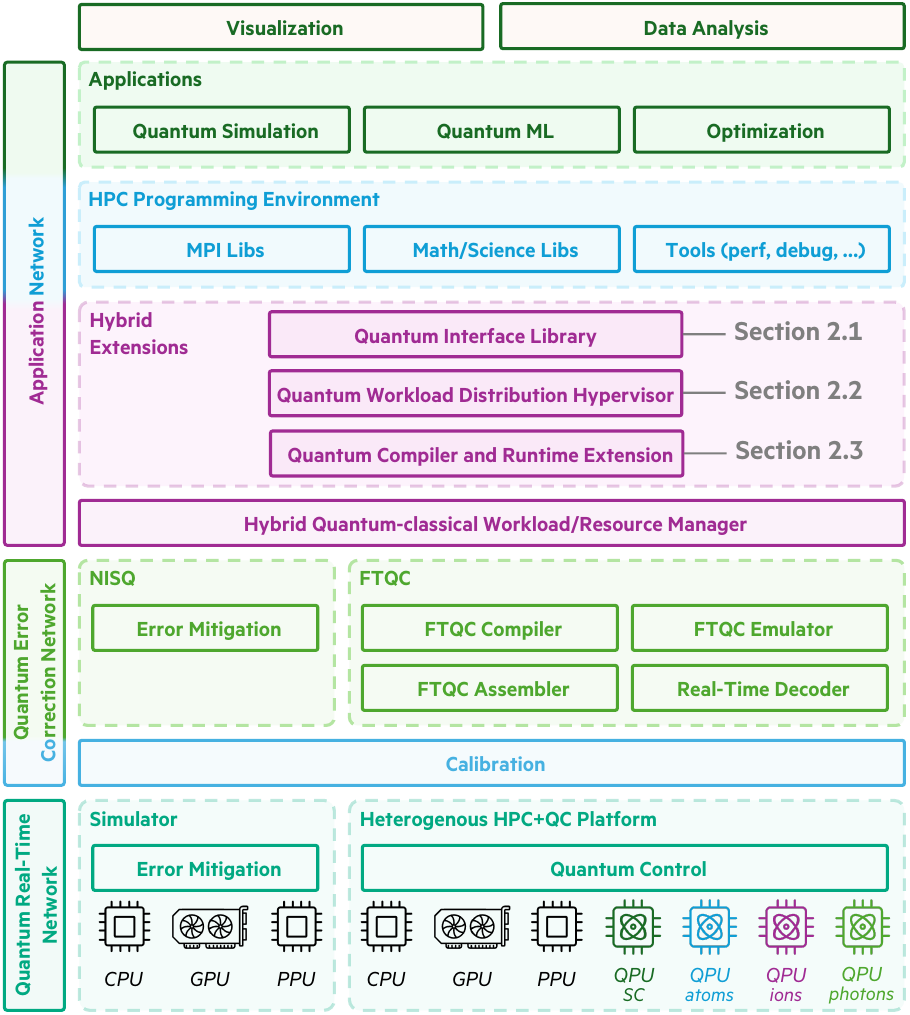}
	\caption{A full-stack architecture highlighting extensions to the HPC programming 
environment for a quantum API, workload distribution hypervisor (ACK), and quantum 
runtime compilation.}
	\label{fig1}
\end{figure}

A robust quantum classical full stack solution is the key to enable scalable, distributed, and hybrid HPC-QC
workload development. With diverse quantum hardware such as superconducting qubits, trapped ions, neutral atoms, and photonic qubits \cite{suc, tri}, various quantum software packages focus on different aspects of quantum computing, e.g. circuit synthesis or optimizing complex design processes including qubit allocation, auxiliary qubit reuse, error mitigation, and quantum error correction. The majority of these software packages adopt declarative frameworks—Python-based Domain-Specific-Languages (DSL)—which facilitate fast experimentation and have built-in REST client support \cite{dec}. When scaling to large circuit sizes with $\sim$100-qubits or greater, declarative frameworks struggle to handle compilation bottlenecks and latency between classical and quantum components in the program, even in the current era of Noisy Intermediate Scale Quantum (NISQ) systems \cite{nisq}. Leveraging classical compilation tool chains such as Clang/LLVM is crucial for developing a large-scale hybrid HPC-QC workload with research efforts moving in this direction \cite{llvmq, plq}. In addition, the vast majority of science and math libraries that support massively parallel distributed computing used to accelerate large-scale classical HPC applications adopts compiled languages such as C, C++, and Fortran for optimal performance.  Hence, integrating quantum kernels within HPC applications will need to take this into consideration. For scalable quantum computing in both the NISQ era (to study systems larger than $\sim$100-qubits) and future fault-tolerant era (to parallelize quantum error correction), efficient partitioning and distribution of large quantum workloads across QPUs for parallel execution is needed. With a good understanding of these challenges and the diversity of current quantum software landscape, we are addressing three main aspects in our design: comparability, performance and scalability. 

\section{Hybrid HPC-QC Full Stack Architecture}

We adopt a modular hardware/device-agnostic approach toward the development of a full quantum-classical stack. We decompose the QC extensions to the HPC programming environment into three tracks: \textbf{1)} a quantum application programming interface library extension, \textbf{2)} an adaptive circuit knitting hypervisor for quantum workload distribution, and \textbf{3)} quantum compilation and runtime extension. Figure ~\ref{fig1} demonstrates the hierarchical architecture diagram of the proposed quantum-classical full stack solution with all three hybrid extension components highlighted with their corresponding description sections.

At the highest level of the stack is hybrid application development, including but not limited to quantum many body physics and chemistry simulation, quantum machine learning, and optimization. We develop a set of hybrid HPC-QC applications of varying algorithmic complexity and scale covering these different areas in order to demonstrate functionality and validate execution viability for all three extensions. These applications are discussed in more detail in the sections below.

\vspace{3mm}

We developed a quantum interface library to enable seamless invocation of quantum kernels from vendor-specific quantum SDKs within HPC applications developed in C/C++ and Fortran. Circuit simulation is replaced with quantum API calls. With this approach, minimal changes to the programming and compilation of massively parallel classical HPC applications (combined with science and math libraries) are needed for application developers, facilitating experimentation with different quantum SDKs. The quantum interface library is discussed in detail in Section ~\ref{sec:qil} below.

In parallel with the quantum API extension, a novel adaptive circuit knitting (ACK) toolkit is being developed as a hypervisor for quantum workload distribution over multi-QPU systems. The ACK algorithm finds efficient partitions of quantum circuits as they evolve by cutting the circuits (in space and time) in locations that minimize entanglement between partitions, overcoming the exponential classical post-processing of standard circuit knitting. It is utilized to manage classical and quantum communication between compute nodes tasked with sub-circuit execution and measurement reconstruction, dispatching large-scale quantum system simulation or quantum machine learning workloads onto smaller QPUs for parallel execution. Section ~\ref{sec:ack} describes the ACK method in more detail and includes simulation demonstrations on 40-qubit spin systems.

The lowest level of the hybrid extensions includes quantum compiler and runtime extensions that are under development to enable performant language-level support for quantum constructs as well as to address the bottlenecks in compile-time with increasing circuit size. These extensions are discussed in Section ~\ref{sec:compiler_ext}.

As shown in Figure ~\ref{fig1}, the components of the full stack below  hybrid workload manager enables quantum error correction/mitigation, calibration, control and execution on simulators as well as the heterogeneous HPC+QC hardware platform. Detailed descriptions of these components can be found in Ref \cite{mohseni2024build}.

With our device-agnostic, heterogeneous framework, the hardware platform can incorporate different types of QPUs, e.g., superconducting, neutral atom, trapped ion, or photonic qubits. It should be mentioned that integrating heterogeneous QPU modalities faces various levels of technology challenges, especially if the opportunity for adaptive circuit knitting does not exist and high-quality quantum interconnects become necessary. Due to differences in QPU modalities---clock frequency, graph conductivities, noise models, ease of long-range interaction, etc.---there will be significant challenges in orchestration of quantum error correction, choosing optimal codes, and optimal real-time decoding among heterogeneous QPUs. We are addressing these challenges as part of the development efforts of our hybrid quantum-classical full stack.

\subsection{Quantum Interface Library}
\label{sec:qil}
The quantum interface library acts as a bridge between traditional HPC applications developed in C/C++ and Fortran and advanced quantum algorithms provided in Python-based DSL, facilitating the integration of quantum computing within existing classical workloads. It provides callable Application Programming Interfaces (APIs) that allow users to access popular quantum algorithms implemented by various quantum SDKs with different programming models, ensuring that HPC applications can leverage quantum resources without requiring fundamental changes to their existing codebase.

\subsubsection{Design of Quantum interface library}
\label{sec:subsubsection}

Quantum Interface Library is designed to expose a set of generic, high-level APIs that abstract away the specifics of quantum SDKs. This design enables the main application logic to invoke quantum kernels without being tightly coupled to the underlying SDK-specific interfaces, functions, or runtime implementations. It also enables hybrid workloads to be submitted to a variety of quantum hardware platforms and simulators through supported SDKs. Different SDKs may interface with multiple types of physical QPUs, such as superconducting qubits, trapped ions, etc. 

This design ensures portability across heterogeneous quantum platforms by allowing a single quantum-classical application to execute on different quantum devices with minimal or no code changes. It also promotes extensibility, enabling the framework to integrate emerging quantum SDKs and backends with minimal disruption to the user-facing programming model. By leveraging these abstractions, the current framework can support rapid prototyping and deployment of hybrid quantum-classical workloads, while remaining adaptable to the evolving quantum software and hardware landscape.

\begin{figure}[!htbp]
\centerline{{\includegraphics[width=1.1\columnwidth]{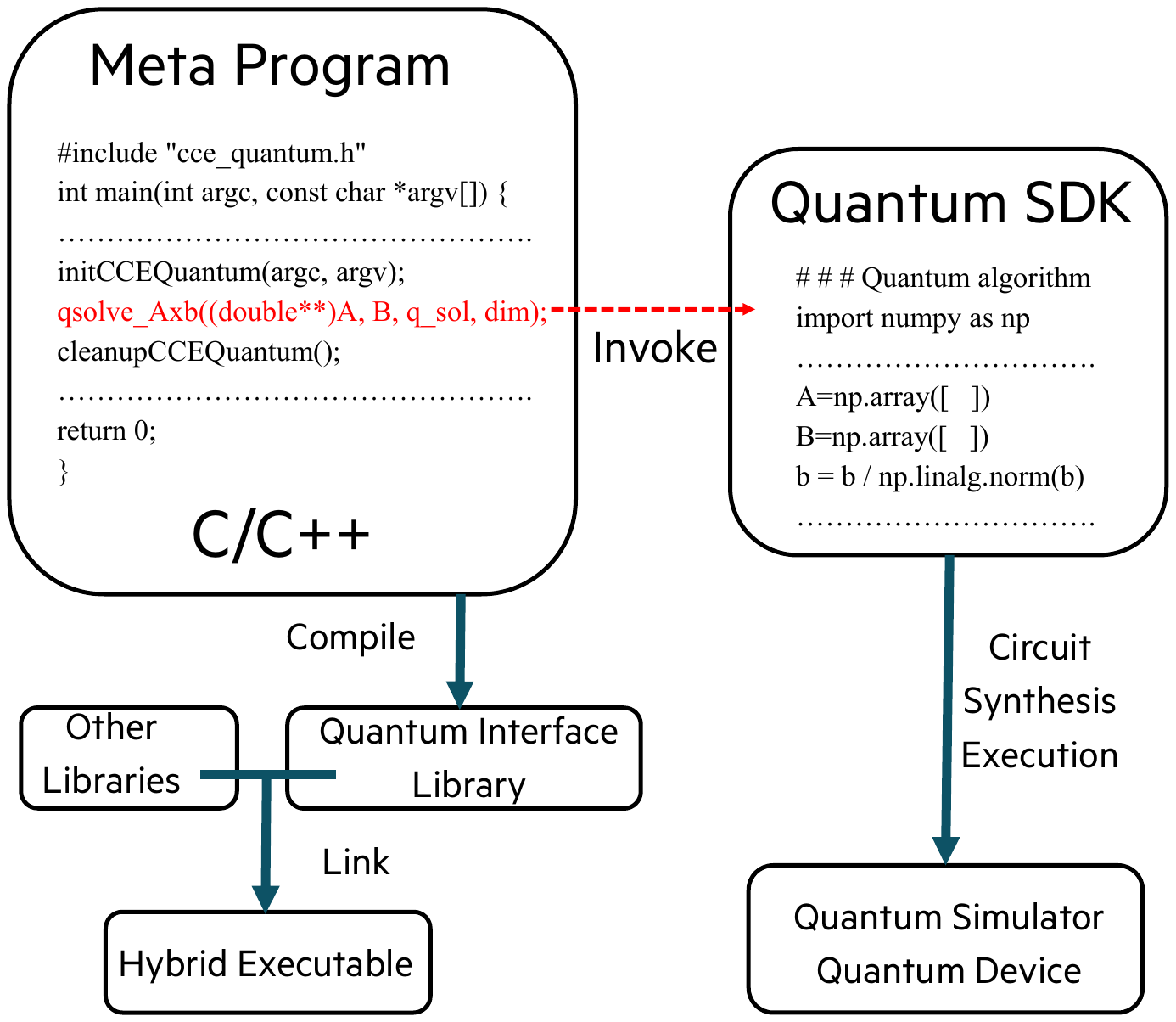}}}
\caption{ Quantum interface library replaces circuit simulation with API calls to enable invocation of quantum kernel from classical HPC program. }
\label{fig2}
\end{figure}

The interface library itself is implemented in C, providing low-level, efficient, and stable APIs that are compatible with C/C++ and Fortran applications. Rather than embedding the quantum logic directly into the library, the interface library routes quantum API calls to/from quantum SDKs within the application meta-program, with appropriate data handling and transfer during runtime. HPC applications can continue to be built with classical CPE C/C++ and Fortran compilers and linked with other libraries such as Cray Science and Math Libraries (CSML)\cite{CSML} as well as the quantum interface library, as illustrated in Figure ~\ref{fig1}. Dynamic linking allows quantum SDKs to be updated independently without recompiling the interface library or hybrid application. The interface library calls vendor-specific quantum SDK routines to compile quantum circuits into standard quantum assembly languages such as OpenQASM \cite{qasm} and subsequently execute on various supported gate-based quantum hardware and simulators hosted remotely on cloud or on-premise.

A hybrid MPI execution model is adopted wherein classical computation and quantum simulation are executed as separate MPI processes but communicate via MPI messaging. The classical process is responsible for handling classical computations, setting up and invoking the quantum algorithm, and sending inputs to and receiving outputs from the quantum simulation. Quantum process is responsible for circuit synthesis and execution. Using MPI to separate these tasks, as shown in Figure~\ref{fig2}, allows:
\begin{itemize}[leftmargin=*, itemsep=0.3em]
\item {\textit{}{Efficient parallel execution for performance}}: Quantum processing can be offloaded to the simulator or hardware device asynchronously. 
\item {\textit{}{Scalability}}: Classical and quantum processes can be expanded independently across multiple CPU, GPU, and QPU nodes.
\item {\textit{}{Flexibility and modularity}}: Keeping the classical and quantum parts separate allows easier debugging and swapping out different quantum backends.
\end{itemize}

\begin{figure}[!htbp]
\centerline{{\includegraphics[width=1.1\columnwidth]{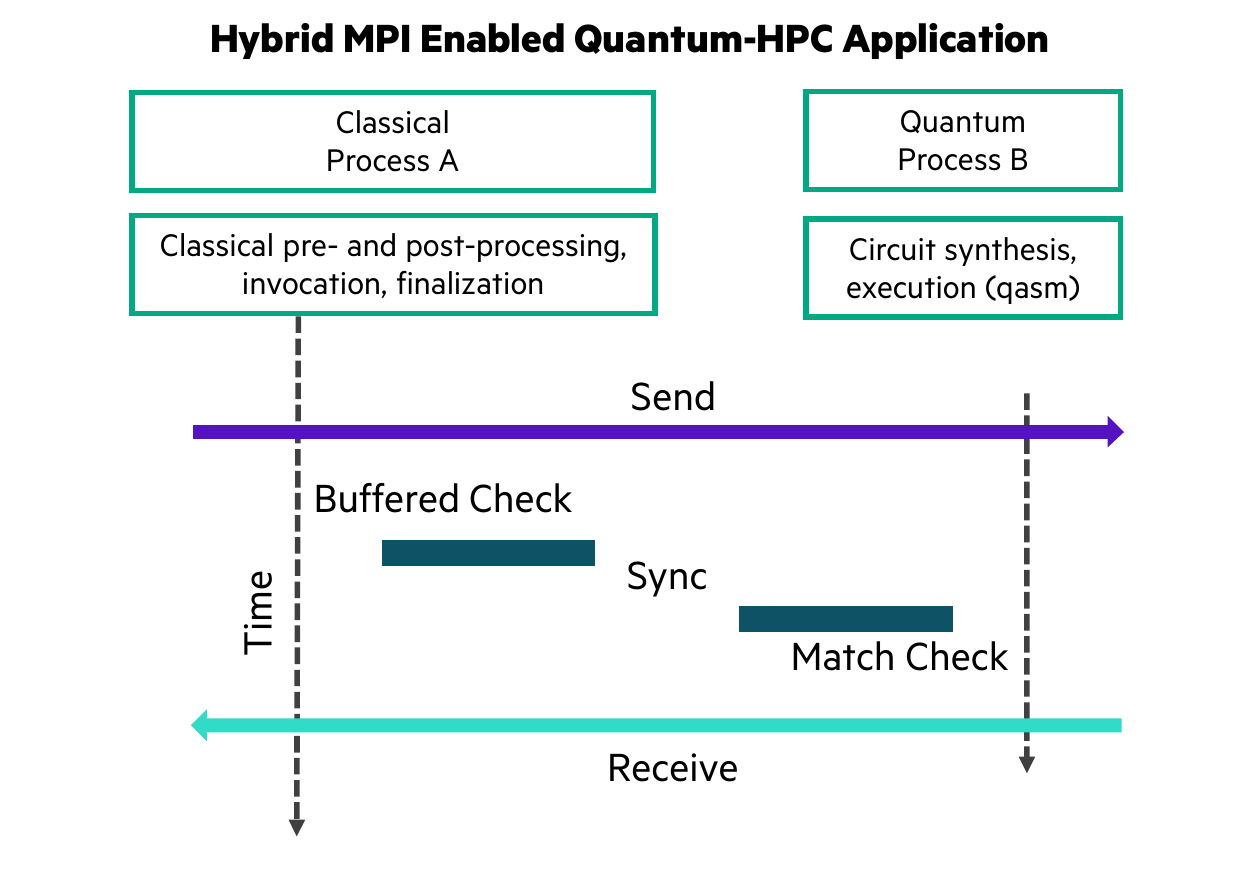}}}
\caption{ Hybrid MPI enabled Quantum-HPC application scheme.}
\label{fig2}
\end{figure}

\begin{figure}[!htbp]
\centerline{{\includegraphics[width=0.9\columnwidth]{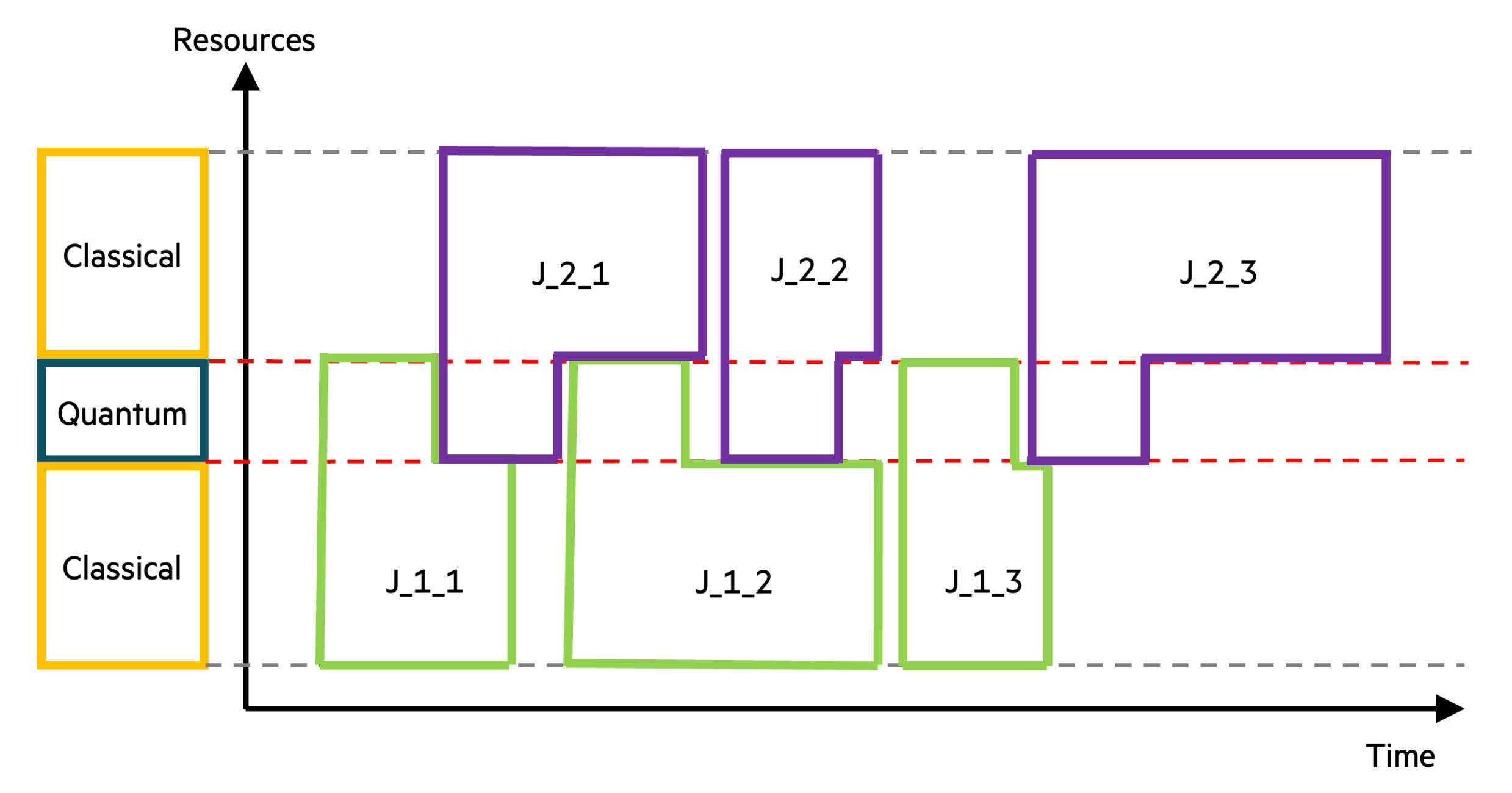}}}
\caption{Monolithic hybrid jobs are split into basic blocks for optimized SLURM scheduling. The indices in \texttt{J\_i\_j} indicate the affiliation to the original job \texttt{i} and time ordering \texttt{j}. The classical resources are typically larger than the quantum resources. }
\label{fig:optimal_scheduling}
\end{figure}

Hybrid HPC-QC job scheduling are configured and implemented with SLURM workload manager and dynamic process management in MPI for a client-server infrastructure \cite{esposito2025slurmheterogeneousjobshybrid}. Monolithic hybrid jobs with large intervals between quantum computation phases, would cause significant idle time of the quantum resources. Splitting these jobs in basic building blocks and using SLURM dependencies opens opportunities to significantly reduce the idle time of the quantum device as show in Figure~\ref{fig:optimal_scheduling}. Users can submit quantum circuits and observables and retrieve a distribution or expectation value later on, while being able to do to classical computations asynchronously. Developed quantum programming interface library is utilized to deliver two MPI enabled hybrid HPC-QC applications in both Python and
C/C++ described in the following two sub-sections. Quantum kernels for circuit synthesis and execution are provided by Classiq’s Python based quantum SDK \cite{Classiq}.

\subsubsection{Case A: Quantum linear solver}
\label{sec:subsubsection}
The first use case is to solve linear systems of equations with the Harrow-Hassidim-Lloyd (HHL) algorithm\cite{hhl}. HHL algorithm is a basic quantum algorithm for solving a set of linear equations: \begin{math}
A\vec{x} = \vec{b}
\end{math}, represented by an \begin{math}
N\times N
\end{math} matrix \begin{math}
A
\end{math} and a vector \begin{math}
\vec{b}
\end{math} of size \begin{math}
N = 2^n
\end{math}.  The solution to the problem is designated by variable  \begin{math}
\vec{x}
\end{math}. In the quantum setting, we reformulate this task by encoding \( \vec{b} \) as a normalized quantum state \( |b\rangle \in \mathbb{C}^N \), and seek to prepare a quantum state \( |x\rangle \propto A^{-1} |b\rangle \) that is proportional to the classical solution \( \vec{x} \), i.e.,\[|x\rangle \propto A^{-1} |b\rangle\]

The workflow for implementing HHL involves classical pre-processing, quantum circuit synthesis, execution, and result validation. The classical pre-processing we need to perform is to decompose matrix $A$ into a sum of Pauli Strings. The efficiency of decomposition can vary depending on the choice of naive or optimized decomposition and the number of Pauli terms, which is  dependent on the structure of matrix $A$. Here we use a naive decomposition of matrix $A$ into 10 Pauli strings, where each matrix entry contributes independently to the Pauli expansion, as shown in the run log in Figure 4.  

The algorithm estimates a quantum state from the initial quantum state corresponding to n-dimensional vector.
Estimation is done by inverting eigenvalues encoded in phases of eigenstates with use of Quantum Phase Estimation (QPE) and amplitude encoding. Now that we have finished with pre-processing we can turn to defining the HHL algorithm. The algorithm consists of 4 steps:
1) State preparation of the RHS vector $\vec{b}$.
2) QPE for the unitary matrix $e^{2\pi iA}$, which encodes eigenvalues on a quantum register of size $m$.
3) An inversion algorithm which loads amplitudes according to the inverse of the eigenvalue registers.
4) An inverse QPE with the parameters in (2). The resulting synthesized circuit is executed on state vector simulator by   $Qiskit$ $Aer$ \cite{Qiskit} in the interest of obtaining an exact solution. Hybrid executable can be submitted and scheduled for parallel execution as regular HPC workloads by a workload manager such as Slurm or PBS as shown in the bottom image of Figure 5.
\begingroup
\setlength{\intextsep}{4pt}        
\setlength{\textfloatsep}{4pt}     
\setlength{\abovecaptionskip}{4pt} 
\setlength{\belowcaptionskip}{10pt} 

\vspace{-6pt}                      
\begin{figure}[!htbp]
\centering
\centerline{{\includegraphics[width=1\columnwidth]{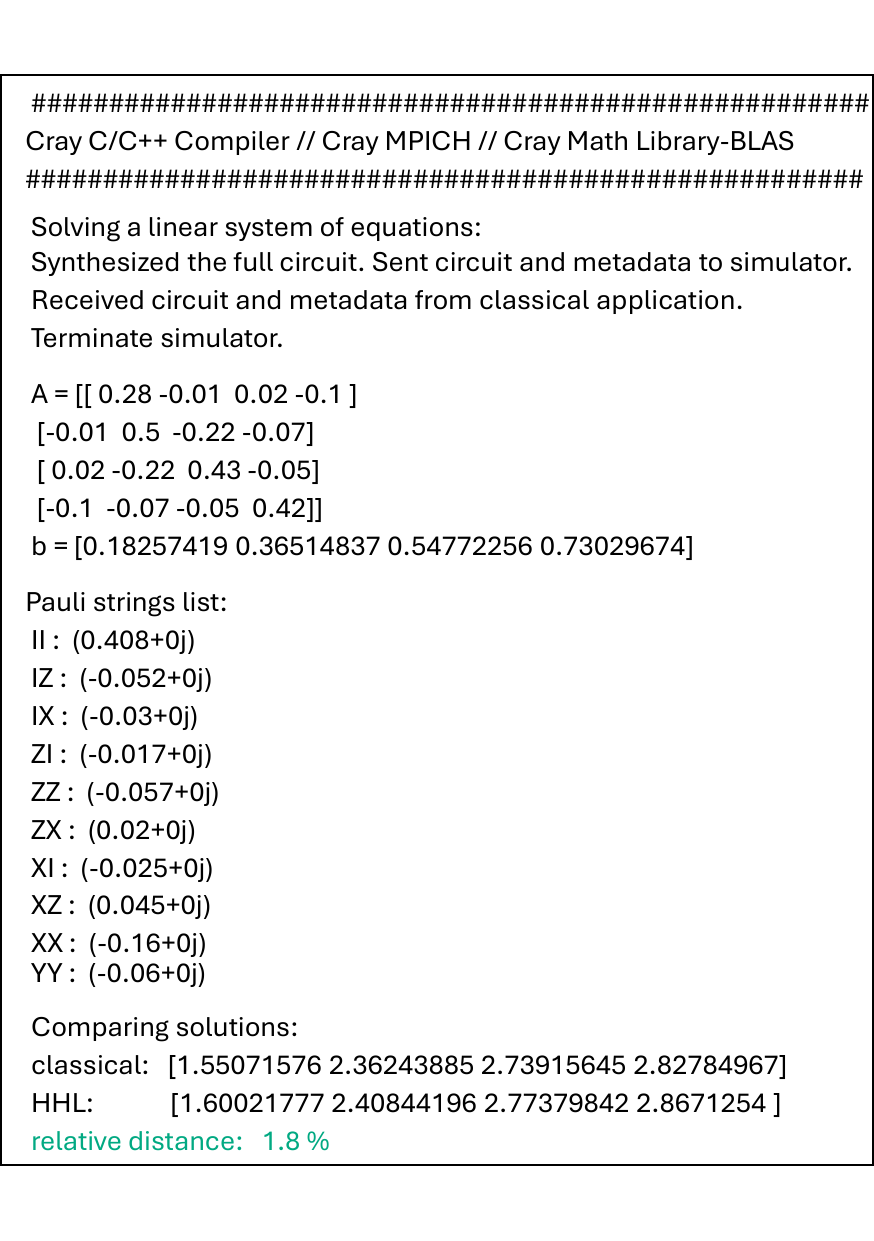}}}
\vspace{-8pt}
\centerline{\boxed{\includegraphics[width=1\columnwidth]{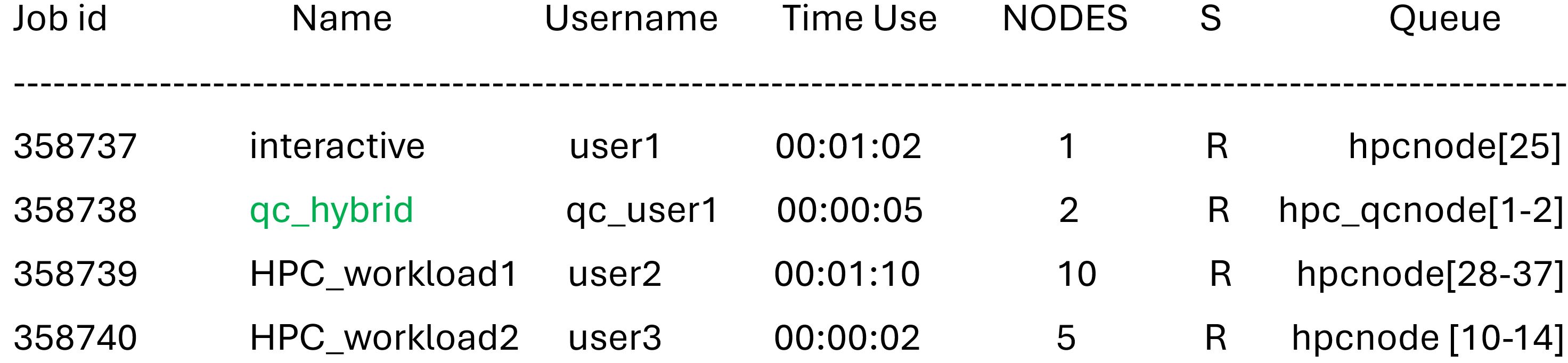}}}
\caption{Hybrid quantum linear solver run log and results comparison with classical solution from BLAS (top). Schematic SLURM job status query (bottom).}
\label{fig4}
\end{figure}
\endgroup

The final quantum state must be processed to obtain a classical solution for comparison. Post-processing is necessary to reconstruct the classical solution by computing the expectation values of Pauli measurements and normalizing. 
The HHL solution is compared with the solution from the BLAS (Basic Linear Algebra Subroutines) library provided in Cray’s scientific and mathematics library, and demonstrated less than 2\% deviation for a 4x4 matrix, as shown in Figure 5. This is a small hybrid workload tested up to a 64x64 matrix on a HPE Cray EX system on two nodes with two AMD EPYC 7763 (Milan) CPUs per node. The primary objective of this first test case is to validate the functional correctness of the quantum interface library and establish the integrated workflow for a hybrid quantum-HPC workload.

\subsubsection{Case B: Quantum optimization}
\label{sec:subsubsection}

The second workload consists of solving MaxCut problems with the quantum approximate optimization algorithm (QAOA) \cite{QAOA}.  Synthesized circuits are provided in gate-level quantum assembly language OpenQASM3 with more complex control flows and parametric circuits. Such circuits are needed for variational algorithms like QAOA, which rely on quantum circuits (ansatz) parameterized by real-valued parameters and classical optimization that updates these parameters iteratively. 

QAOA stands out as a prominent hybrid quantum-classical technique for addressing challenging combinatorial optimization problems, including the MaxCut problem, which involves finding the best partition of a graph's vertices into two sets to maximize the number of edges between them. The QAOA variational quantum state is given by:
\begin{displaymath}
|\psi(\boldsymbol{\gamma}, \boldsymbol{\beta})\rangle = \prod_{j=1}^p e^{-i \beta_j H_M} e^{-i \gamma_j H_C} |+\rangle^{\otimes n}
\end{displaymath}

where
\begin{itemize}
  \item \( |\psi(\boldsymbol{\gamma}, \boldsymbol{\beta})\rangle \) is the QAOA ansatz after \( p \) layers,
  \item \( \boldsymbol{\gamma} = (\gamma_1, \dots, \gamma_p) \) and \( \boldsymbol{\beta} = (\beta_1, \dots, \beta_p) \) are variational parameters,
  \item \( H_C \) is the cost Hamiltonian encoding the optimization problem,
  \item \( H_M = \sum_{i=1}^n X_i \) is the mixer Hamiltonian (sum of Pauli-\(X\) operators),
  \item \( |+\rangle^{\otimes n} \) is the initial equal superposition state.
\end{itemize}

The goal is to find the optimal parameters \( \boldsymbol{\gamma} \) and \( \boldsymbol{\beta} \) that maximize the expected cost:
\begin{displaymath}
\max_{\boldsymbol{\gamma}, \boldsymbol{\beta}} 
\; \langle \psi(\boldsymbol{\gamma}, \boldsymbol{\beta}) | 
H_C | \psi(\boldsymbol{\gamma}, \boldsymbol{\beta}) \rangle .
\end{displaymath}

\sloppy
To scale this approach for large MaxCut instances 
(e.g., 500–10{,}000 nodes), we explore the QAOA-in-QAOA 
(\texttt{QAOA$^2$})~\cite{QAOA2} framework in this study. 
\texttt{QAOA$^2$} employs a divide-and-conquer strategy: the original 
graph is decomposed into smaller subgraphs, each of which can be 
independently solved—often in parallel—using QAOA, thereby enabling 
efficient utilization of multiple quantum resources. 
This approach is particularly useful for current NISQ devices with a 
limited number of qubits (\(\sim\!100\) qubits).

The hybrid quantum-classical MaxCut workflow using QAOA2 is illustrated in Figure 6 with results collected and compared on classical rank 0. The approach partitions a large input graph into smaller subgraphs that can be mapped to NISQ-capable circuits. The classical graph partitioning and community-level Maxcut merging are managed by classical HPC process, while quantum simulation of subgraphs is simulated using distributed memory parallelism over MPI. Each quantum process simulates a quantum device and performs QAOA optimization on its assigned subgraph. Results are sent back to rank 0 and gathered asynchronously, and subgraph results are combined into a unified solution. The QAOA results are benchmarked against classical approaches (Goemans-Williamson, Greedy, Random) for node counts up to 2500 using up to 512 CPU nodes on the same HPE Cray EX system as described in the previous case.  Greedy and GW methods can still yield strong results, especially on small or moderately sized graphs. QAOA gives a reasonable approximation and performs better on larger structured graphs, as expected. This application was demonstrated live at ISC24 using a 20-qubit IQM quantum device accessed from the LUMI supercomputer in Finland \cite{maxcutIQM}. 

\begingroup
\setlength{\intextsep}{4pt}        
\setlength{\textfloatsep}{4pt}     
\setlength{\abovecaptionskip}{-16pt} 
\setlength{\belowcaptionskip}{0pt}

\vspace{-6pt}                      
\begin{figure}[H]                  
\centering
\includegraphics[width=\columnwidth]{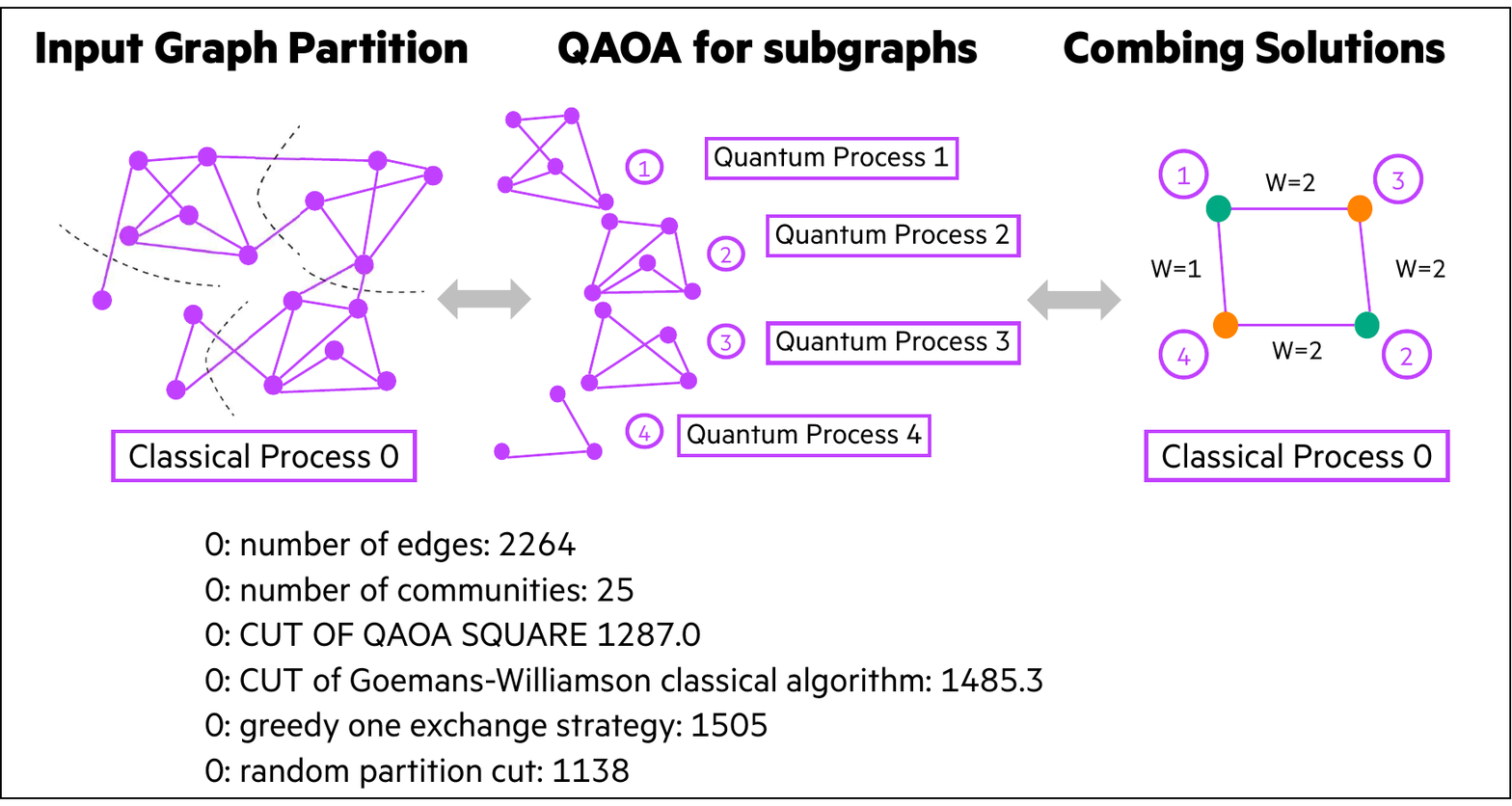}
\caption{Hybrid distributed Quantum MaxCut with QAOA2 scheme (top) and results comparison with classical algorithms (bottom).}
\label{fig3}
\end{figure}
\vspace{-8pt}                      
\endgroup
\subsection{Adaptive Circuit Knitting Hypervisor}
\label{sec:ack}

For quantum computing to operate at scale, it will be necessary to efficiently parallelize over many QPUs, possibly of different modalities, integrating them as co-processors within an HPC framework. Quantum interconnects between QPUs may not be available in the near term. We will need to rely on classical HPC interconnects.

For hybrid quantum-classical computing to scale, efficient partitioning and distribution of quantum workloads across quantum processing units (QPUs) is crucial. To address the scalability of QC, a novel adaptive circuit knitting (ACK) toolkit has been developed with initial testing conducted on 1D quantum spin chains.

ACK can efficiently partition a larger quantum circuit into smaller sub-circuits by finding the optimal cuts locations of minimal entanglement on quantum gates between qubits. 
 Especially,  our method decreases the sampling overhead of circuit knitting by cutting gates in locations that minimize entanglement between partitions. 

Distributing quantum computation is a highly non-trivial task and presents fundamentally different challenges and opportunities compared to classical HPC parallelization models. This is primarily due to the presence of entanglement, a non-local quantum correlation between qubits that cannot be cleanly partitioned in space or logic. Entanglement is essential to quantum advantage but also introduces strong coupling between distant parts of a quantum circuit. This makes naive partitioning ineffective: if two highly entangled qubits are placed on different QPUs, then inter-device communication or post-processing is required to reconstruct their joint behavior. Complicating this further, the structure of entanglement can be dynamic and problem-dependent — it is often not known in advance which qubits are most entangled and where circuit cuts can be made with minimal loss of quantum coherence. To address this, the circuit knitting technique \cite{circuit_kniting} has been recently developed, where large quantum circuits are split into subcircuits, executed independently, and their observables recombined via classical post-processing. While this allows for some form of parallel execution and even distribution across classical HPC nodes, it incurs an exponential sampling overhead if entanglement between partitions is high.  

\begin{figure}[H]
    \centering
    \setlength{\fboxrule}{0.6pt}    
    \setlength{\fboxsep}{1pt}
    \fbox{\includegraphics[width=0.985\columnwidth]{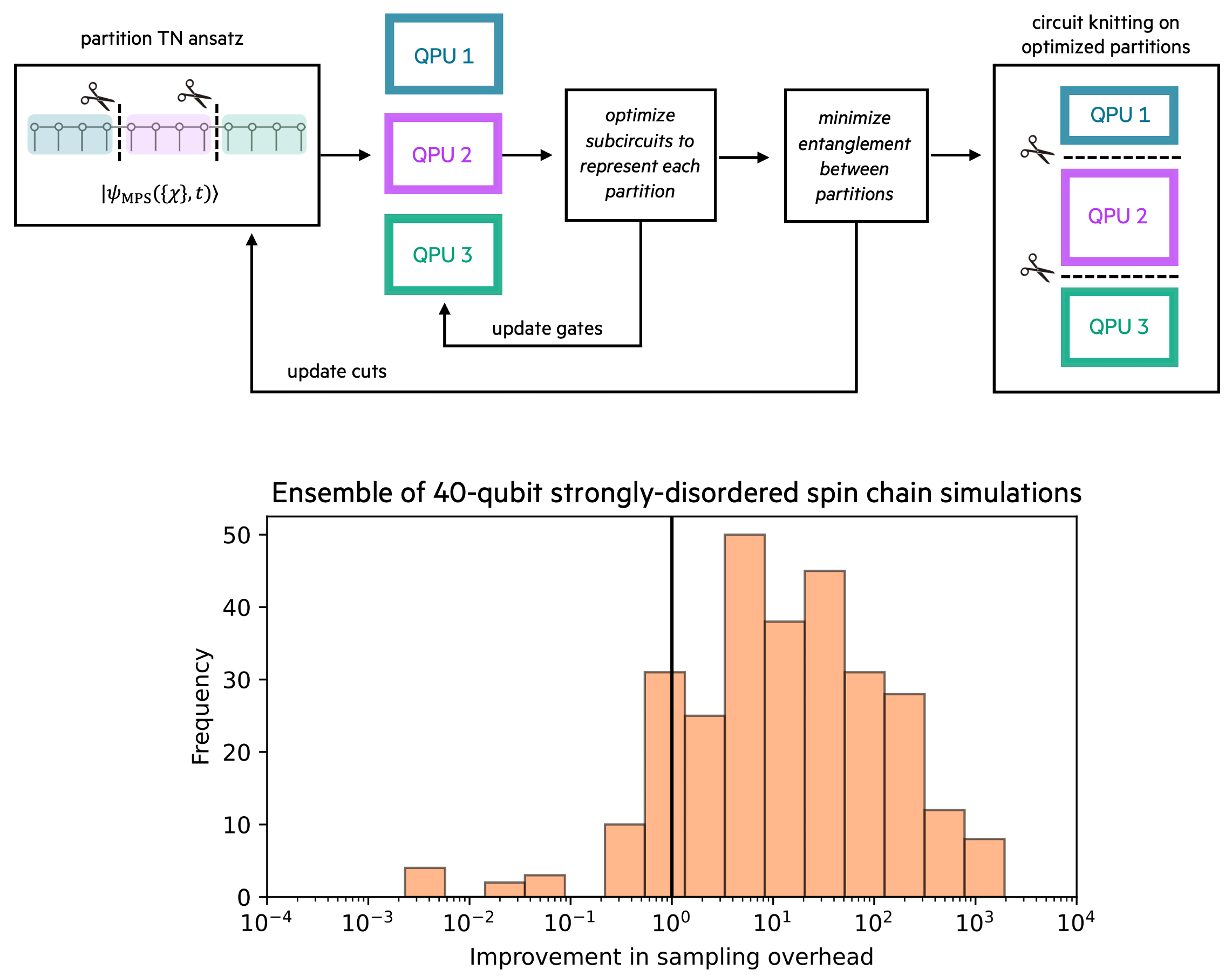}}
    \caption{Adaptive circuit knitting hypervisor scheme (top) and 
    sampling overhead reduction for 40-qubit strongly disordered spin 
    chains (bottom).}
    \label{fig5}
    \vspace{-6pt}
\end{figure}

To mitigate this cost, a more sophisticated  adaptive circuit knitting method has been developed. In this approach, the quantum circuit is analyzed — and partitioned — in a way that explicitly minimizes the entanglement between subcircuits. One enabling tool is the use of tensor networks (TNs), such as matrix product states (MPS), which represent quantum states in a compressed form that naturally encodes the entanglement structure. These representations allow for an adaptive algorithm to select optimal cut points that minimize entanglement entropy, reducing the number of samples needed to reconstruct observables. As shown in the top panel of Figure ~\ref{fig5}, the process involves an inner loop where subcircuits (inspired by TN structure) are optimized in parallel and an outer loop where the cutting strategy is refined based on entanglement minimization. This fusion of quantum simulation with classical optimization mirrors how GPUs accelerate linear algebra, suggesting a future where QPUs act as accelerators for classically structured quantum workloads.

We demonstrate the potential of this method by applying adaptive circuit knitting to simulate strongly disordered 1D spin chains evolving under an Ising Hamiltonian with both transverse and longitudinal fields. Systems like these are of practical interest in condensed matter physics, and are known to exhibit complex quantum phenomena such as many-body localization, which can be difficult to simulate classically. Using an entropy-guided cutting strategy, circuits of up to 32 qubits can be simulated on a single node with 4 A100 GPUs with 40GB HBM2 DRAM. For the 40-qubit simulations, 256 nodes on Perlmutter with 4 A100 GPUs each were utilized using CUDA-Q SDK with GPU-accelerated quantum simulator backends \cite{Mohseni_gtc_talk} in collaboration with NVIDIA. CUDA-Q \cite{cuda_q} runtime manages execution, ensuring integration across CPUs, GPUs, and QPUs. CUDA-Q compilation targets include multi-GPU simulation backends (e.g., statevector, density matrix, tensor networks) and physical QPUs (e.g., Quantinuum, IonQ, IQM, OQC). We observed significant reductions in sampling overhead compared to baseline load-balanced cuts — most cases with 10-100x improvement and for a few cases over 1000x improvement, as shown in the lower panel of Figure ~\ref{fig5}. 

\subsection{Quantum Compiler Extension}
\label{sec:compiler_ext}

At the lower level, the quantum compiler and runtime extension are being 
developed to enable performant, language-level support for quantum constructs 
and quantum co-processor programming capabilities. Compatibility with existing 
classical compilers, libraries, and utilities that perform code analysis during 
compilation—and automatically generate highly optimized code—can be achieved 
through direct integration. The main goal is to address the compile-time 
bottleneck that arises with increasing circuit size (larger gate counts, circuit 
depth, and qubit numbers) and to minimize latency between classical and quantum 
components within the same program.

The framework of quantum compiler and runtime extensions is grounded in the following design principles: 

\begin{itemize}[leftmargin=*, itemsep=0.3em]

\item \textit{Integrated Quantum Co-Processor Programming Model:}
Quantum Processing Units (QPUs) are integrated as accelerators alongside CPUs 
and GPUs within a heterogeneous HPC system. Rather than treating this as a 
runtime coordination of a single application workflow, the model enables 
compilation of hybrid applications that offload selected quantum kernels to 
QPUs. These kernels may include quantum circuit synthesis, parametric 
\emph{ansatz} generation, or iterative quantum–classical routines.

\item \textit{Compiler-Level Interoperability Across Frontends:}
Frontend agnosticism is achieved by extending the LLVM Intermediate 
Representation (IR) with the Quantum Intermediate Representation (QIR). 
Developed by Microsoft, QIR serves as a common interface between multiple 
quantum languages and hardware backends. It allows ingestion of IR emitted 
by a variety of quantum compilers such as CUDA-Q (NVQ++), \texttt{Q\#}, and 
OpenQASM-based toolchains.

\item \textit{Hardware-Retargetable Code Generation and Linking:}
This layer leverages existing classical LLVM compilation frameworks to 
consume LLVM IR and QIR emitted from different quantum compilers and 
frontends. Assembly code for a given target architecture is generated directly 
from the input IR without requiring non-standard extensions. By linking with 
the appropriate runtime libraries provided by hardware vendors, execution on 
different quantum processors can be achieved in a quantum-backend-retargetable 
manner.

\end{itemize}

\setlength{\textfloatsep}{12pt}
\setlength{\abovecaptionskip}{6pt}
\setlength{\belowcaptionskip}{0pt}
\vspace{-8pt}

\begin{figure}[!htbp]
\centering
\begin{minipage}{0.98\columnwidth}
\lstset{
  frame=single,
  captionpos=b,
  basicstyle=\ttfamily\scriptsize,
  aboveskip=1pt,
  belowskip=1pt,
  xleftmargin=0.5em,
  framexleftmargin=0.5em,
  showstringspaces=false,
  keepspaces=true,
  breaklines=true,
  rulecolor=\color{black},
  keywordstyle=\color{blue}\bfseries,
  commentstyle=\color{gray}\itshape,
  emph={declare,call,br,label,phi,icmp},
  emphstyle=\color{blue},
  moredelim=[l][\color{gray}]{\#}, 
}

\begin{lstlisting}
# QIR types and intrinsics
%Array  = type opaque
%Qubit  = type opaque
%Result = type opaque

declare %Array*  @__quantum__rt__qubit_allocate_array(i64)
declare void     @__quantum__qis__h(%Qubit*)
declare void     @__quantum__qis__x__ctl(%Array*, %Qubit*)
declare %Result* @__quantum__qis__mz(%Qubit*)

# GHZ kernel (partial)
entry:
  %qvec = call %Array* @__quantum__rt__qubit_allocate_array(i64 %n_qubits)
  %zero_idx = call i8* @__quantum__rt__array_get_element_ptr_1d(%Array* %qvec, i64 0)
  %q0 = bitcast i8* %zero_idx to %Qubit*
  call void @__quantum__qis__h(%Qubit* %q0)
  br label %loop

loop:
  %i = phi i64 [0, %entry], [%next_i, %loop_body]
  %cond = icmp ult i64 %i, %n_qubits_minus1
  br i1 %cond, label %loop_body, label %measure
\end{lstlisting}
\end{minipage}
\caption{Partial GHZ kernel in QIR (LLVM IR format).}
\label{fig:ghz-ir}
\end{figure}

Figure~\ref{fig:ghz-ir} shows a partial fragment of the full LLVM IR with QIR emitted from a hybrid quantum-classical program. In a single-source context, a CUDA-Q quantum kernel for the creation and measurement of a 30-qubits GHZ state \cite{GHZ} is embedded in a C++ hybrid application with QIR emitted. LLVM codegen in Cray compiler is utilized to lower input IRs for binary object creation and linked with CUDA-Q runtime libraries for hybrid executable generation, which is executed on a single A100 GPU. It demonstrates how classical control flow, quantum memory management, and quantum instruction emission are interleaved via a unified intermediate representation. Semantic boundaries between quantum and classical computation is retained. 

The modular, extensible design that builds on established Cray compiler infrastructure ensures that as quantum hardware evolves, the same compilation strategy can be retargeted, optimized, and deployed across future heterogeneous systems with QPUs constructed from diverse qubit modalities.

\section{Summary}

We have presented our quantum classical full stack solution and modular hardware/device-agnostic approach to address the unique challenges in pairing quantum computing with modern HPC infrastructure. To enable and facilitate the development of distributed, scalable hybrid HPC-QC workloads, the three main aspects we address are compatibility, performance, and scalability.  The architecture design is defined as the development of extensible interfaces for quantum programming (bridging classical applications and various quantum SDKs), dispatching (sub-circuit partitioning for distributing
quantum workloads across multiple QPUs), and compilation (portable quantum compilation and retargetability) within exsiting HPC programming environment.

At this stage, practical outcomes have been demonstrated on a range of hybrid HPC-QC workloads. Distributed hybrid HPC-QC applications including a quantum liner solver, quantum optimization using QAOA, and 1D spin chain simulation from smaller scale (2 AMD EPYC 7763 (Milan) CPU nodes) to large scale (up to 1024 A100 GPUs across 256 GPU nodes) have been constructed to validate the functionalities of quantum extensions developed by far. Existing infrastructure for HPC can be leveraged for future HPC-Quantum hybrid system in data and user management, process scheduling, control and networking to a large extent. Building on top of existing HPC toolchain with broad architecture support can significantly reduce development time and avoid complexity associated with compatibility issues. To ensure effective implementation and testing of QC extensions under development, we will continue to collaborate with multiple quantum software and hardware partners, and obtain valuable feedback from broader Quantum and HPC communities.

\section{Acknowledgments}

We would like to express our sincere gratitude to our collaborators at NVIDIA (Pooja Rao, Yuri
Alexeev) for insightful discussions on the adaptive circuit knitting algorithm as well as collaboration on implementation and simulation using CUDA-Q. We would like to acknowledge Classiq (Tamuz Danzig) and give a special thanks to HPE Cray Compilation
Environment Team for all the support.
%


\vspace{15pt}
\setlength{\bibsep}{2pt plus 0.3ex}
\bibliography{references}

\end{document}